\documentclass[amsmath,amssymb,prl,reprint]{revtex4-1}
\usepackage{graphicx}
\usepackage{dcolumn}
\usepackage{bm}

\setcitestyle{super}

\begin{document}

\title{Noise thermometry applied to thermoelectric measurements in InAs nanowires}
\author{E.S.~Tikhonov}
\affiliation{Institute of Solid State Physics, Russian Academy of
Sciences, 142432 Chernogolovka, Russian Federation}
\affiliation{Moscow Institute of Physics and Technology, Dolgoprudny, 141700 Russian Federation}
\author{D.V.~Shovkun}
\affiliation{Institute of Solid State Physics, Russian Academy of
Sciences, 142432 Chernogolovka, Russian Federation}
\affiliation{Moscow Institute of Physics and Technology, Dolgoprudny, 141700 Russian Federation}
\author{D.~Ercolani}
\affiliation{NEST, Istituto Nanoscienze -- CNR and Scuola Normale Superiore, Piazza S. Silvestro 12, I-56127 Pisa, Italy}
\author{F.~Rossella}
\affiliation{NEST, Istituto Nanoscienze -- CNR and Scuola Normale Superiore, Piazza S. Silvestro 12, I-56127 Pisa, Italy}
\author{M.~Rocci}
\affiliation{NEST, Istituto Nanoscienze -- CNR and Scuola Normale Superiore, Piazza S. Silvestro 12, I-56127 Pisa, Italy}
\author{L.~Sorba}
\affiliation{NEST, Istituto Nanoscienze -- CNR and Scuola Normale Superiore, Piazza S. Silvestro 12, I-56127 Pisa, Italy}
\author{S.~Roddaro}
\affiliation{NEST, Istituto Nanoscienze -- CNR and Scuola Normale Superiore, Piazza S. Silvestro 12, I-56127 Pisa, Italy}
\author{V.S.~Khrapai} 
\affiliation{Institute of Solid State Physics, Russian Academy of
Sciences, 142432 Chernogolovka, Russian Federation}
\affiliation{Moscow Institute of Physics and Technology, Dolgoprudny, 141700 Russian Federation}

\begin{abstract} 
We apply  noise thermometry to characterize charge and thermoelectric transport in single InAs nanowires (NWs) at a bath temperature of 4.2 K. Shot noise measurements identify elastic diffusive transport in our NWs with negligible electron-phonon interaction. This enables us to set up a measurement of the diffusion thermopower. Unlike in previous approaches, we make use of a primary electronic noise thermometry to calibrate a thermal bias across the NW. In particular, this enables us to apply a contact heating scheme, which is  much more efficient in creating the thermal bias as compared to conventional substrate heating. The measured thermoelectric Seebeck coefficient exhibits strong mesoscopic fluctuations in dependence on the back-gate voltage that is used to tune the NW carrier density. We analyze the transport and thermoelectric  data in terms of approximate Mott's thermopower relation and to evaluate a gate-voltage to Fermi energy conversion factor. 
\end{abstract}

\maketitle

Efficient thermoelectric (TE) conversion in solid state devices has been an elusive target for many decades. An ideal TE material should display a large electrical conductivity $\sigma$ and Seebeck coefficient $S$, and a small heat conductivity $\kappa$ \cite{Snyder2008}. On the other hand, bulk materials are typically characterized by a strong interdependence between these parameters, which poses limits to the maximum achievable conversion efficiency \cite{Snyder2008}. Nanostructured semiconductors today offer a host of novel ways to elude part of these constraints and are leading to  a promising new direction in TE research \cite{Majumdar2004,Dresselhaus2007,Shi2012}. For instance, present evidences show that phonon conductivity can be significantly suppressed in nanostructures \cite{Boukai2008,Hochbaum2008,Martinez2011,Swinkels2015,Zhou2011} and promising results have also been obtained on the tuning of the TE response through an engineering of electron quantum states \cite{Heremans2008,Tian2012,Wu2013}. The investigation of TE effects in nanoscale conductors, though, brings with it a set of technical challenges linked to reproducibility and accuracy in the estimate of the TE properties of single nanostructures. In particular, electrical and heat contact resistances~\cite{Swinkels2015,Zhou2011} are often difficult to predict and measure, as well as the relative impact of the different transport mechanisms in the emergence of the nanomaterials TE properties. In addition, the role of phonons and of their interaction with the electron system is often hard to access in a real nanostructure \cite{Hoffmann2009}. This calls for novel measurement methods to correlate various aspects of the TE response of a nanomaterial and sort out the fundamental physics ruling their TE behavior \cite{Roddaro2013,Yazji2015}. 

In our work, we investigate the TE response of individual InAs NWs at a temperature of few Kelvins and demonstrate a primary thermometry method based on current noise measurements. Our approach allows first of all a direct measurement of the thermal bias across the device, and covers a fairly large operation temperature range going well beyond the one typically available with superconductive tunnel junctions \cite{Giazotto_Pekola_2006}. In addition, we show that the investigation of shot noise as a function of the bias offers valuable further insight on the device transport regime. In particular, we identify the electron-phonon energy exchange can be neglected for temperatures below  $\sim40$\, K and the transport is consistent with the elastic diffusion regime. This enables us to investigate the diffusion thermopower of the individual InAs NWs in the regime of strong mesoscopic fluctuations.

{\bf Device characterization}

\begin{figure*}[t]
\begin{center}
\vspace{10mm}
  \includegraphics[width=0.8\linewidth]{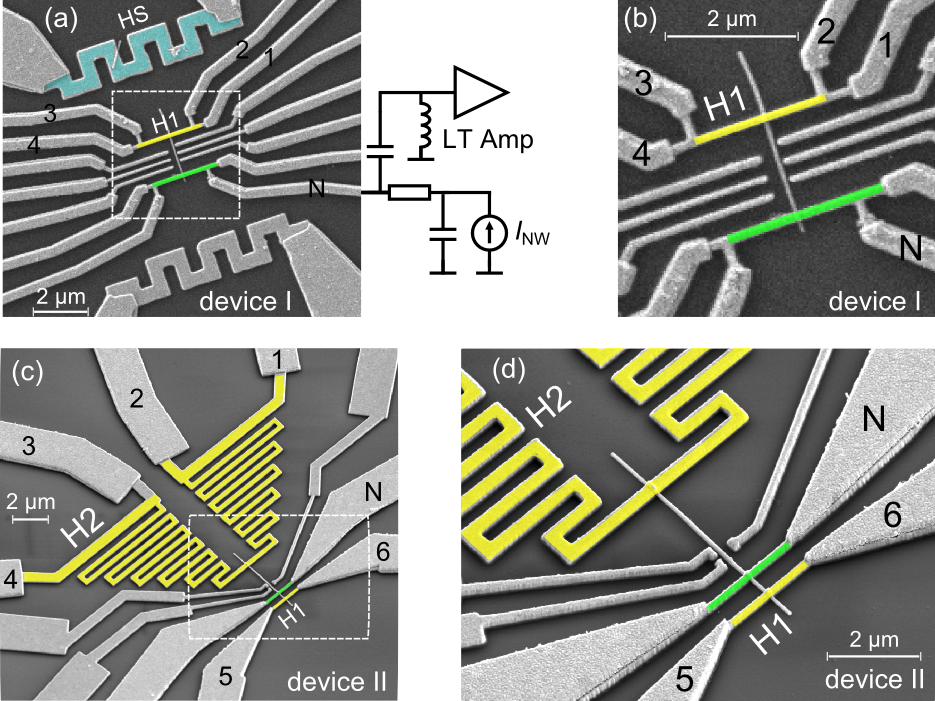}
\end{center}
  \caption{Scanning electron micrographs of two studied devices. (a,b) Device~I. (a) InAs nanowire is connected to two identical $2\,\mathrm{\mu m}$ long Ti/Au metallic contact stripes which further form four-terminal bars. One of the stripes, marked as~N and green colored, is used for the noise measurements, while the other one, marked as~H1 and yellow colored, is used as a contact heater in the TE measurements. A few micrometers above the electrically isolated meander-shaped metallic substrate heater, marked as HS and blue colored, is located. The noise measurement scheme, attached to the terminal  N, is given on the right hand side and is the same for both devices. In the TE measurements we set the heating current~$I_H$ through one of the heaters and measure the resulting voltage~$V_{th}$ between terminals N and 2. Terminal~3 was dc-grounded throughout all the experiments, while terminals 1, 2, 4 and 5 could be either grounded or left floating. All terminals are rf by-passed to ground by $10\,\mathrm{nF}$ capacitors. (b) Magnified inner part of the device. (c,d) Device~II. (c) InAs nanowire is connected to three different contact stripes. Two stripes, marked as~H1 and H2, yellow colored, are used as the contact heaters in the TE measurements, respectively for the short and the long NW sections. The third in-between stripe N, green colored , is used for the noise measurements. In the TE measurements we set the heating current~$I_H$ through one of the heaters and measure the resulting voltage~$V_{th}$ between either terminals (5,6) and N (for the short NW section) or terminals (2,3) and N (for the long NW section). All terminals are  rf by-passed to ground by  $10\,\mathrm{nF}$ capacitors and can be either dc-grounded or left dc-floating during the experiment. (d) Magnified inner part of the device.} 
	\label{fig1}
\end{figure*}

Au-assisted Se doped InAs NWs are grown by chemical beam epitaxy on an InAs(111) B substrate. The NWs of $\approx\,70\,\text{nm}$ diameter and 2$\,\mu$m  length were drop-casted on a doped silicon wafer with 280~nm thick $SiO_2$ insulator on top. The carrier density of the InAs NWs derived by  field effect measurements is about $\rm 1\times10^{18} cm^{-3}$. We performed the measurements in two $^3$He inserts, with the samples immersed in gas (at $T=4.2$\,K). The shot noise spectral density was measured using home-made low-temperature amplifiers (LTamp) with a voltage gain of about 10\,dB, input current noise of $\rm \sim10^{-27}\,A^2/Hz$. We used a resonant tank circuit at the input of the LTamp, see the sketch in Fig.~\ref{fig1}a, with a ground bypass capacitance of a coaxial cable and contact pads $\sim40\,$pF, a hand-wound inductance of $\sim6\,\mu$H and a load resistance of $\rm 10\,k\Omega$. The output of the LTamp was fed into the low noise room temperature amplification stage with a hand-made analogue filter and a power detector. The setup has a bandwidth of $\sim0.5$\,MHz around a  center frequency of $\approx10\,$MHz. A calibration was achieved by means of equilibrium Johnson-Nyquist noise thermometry. For this purpose we used a commercial pHEMT transistor connected in parallel with the device, that was depleted otherwise. All transport measurements were performed with the help of a two-terminal or four-terminal lock-in resistance measurement. 

In our experiments we used four devices of two different architectures shown in Fig.~\ref{fig1}  and referred to as device I and device II below. A larger scale SEM image of the device I is shown in fig.~\ref{fig1}a and the magnified inner part in fig.~\ref{fig1}b. In the figures, the light gray color corresponds to Ti/Au metallic layers evaporated on top of the $\rm SiO_2$ substrate or a single InAs NW. Two contact stripes are used as ohmic contacts to the NW. Each stripe is shaped in the form of a four-terminal bar, whose narrower and thinner part is connected to the either end of the NW. One of them, marked N and greenish, is connected to the dc measurement setup and the low-temperature rf-amplifier via the terminal 1. This contact stripe serves for noise detection. The other one, marked H1 and yellowish, is used as a contact heater. In device I we also used a meander-shaped substrate heater, marked HS and bluewish. Heating currents $I_H$ serve to energize the heaters and create a thermal bias across the NW during the TE measurements. The remaining meander-shaped heater, as well as the plunger next to the NW, were not used and kept grounded in present experiment.  Figs.~\ref{fig1}c and d depict the layout of the device II. This device is equipped with three contact stripes which divide the NW into a short and long section. The center contact, marked N, greenish, has the same meaning as for the device I. The side contact stripes, marked H1 and H2, yellowish,  are used as heaters for the short and the long sections, respectively. The device II is lacking substrate heaters, its plunger gates were also not used in present experiments. In the rest of the paper we discuss the results of measurements obtained in two representative samples of both architectures.

\begin{figure}[t]
\begin{center}
\vspace{10mm}
  \includegraphics[width=0.8\columnwidth]{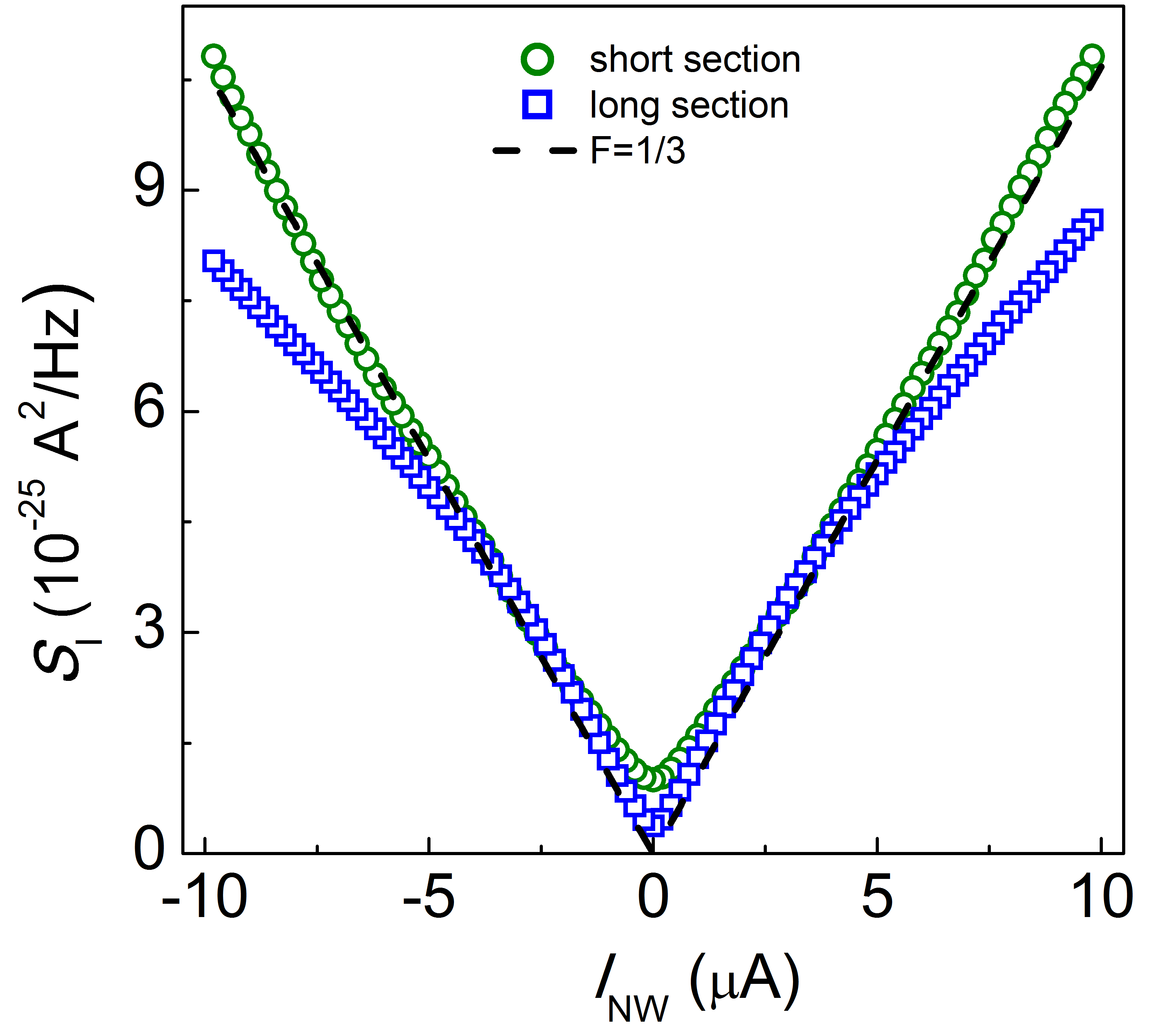}
	   \end{center}
  \caption{Characterization of the transport regime in our NWs. Shot noise spectral density as a function of current at $T=4.2\,\mathrm{K}$ for long (squares) and short (circles) sections of the NW device~II. The slope of the dashed guide line corresponds to the Fano-factor $F=1/3$.}
	\label{fig2}
\end{figure}

In Fig.~\ref{fig2} we characterize the transport regime in our NWs using the shot noise measurements at a bath temperature of $T_0=4.2$~K. Here we plot the noise spectral density $S_I$ for the device II in dependence of the NW bias current $I_{NW}$, which flows via the terminal N to the grounded contact 1 or 3, via, respectively, the short or the long NW section. In both cases, near the origin $S_I$ crosses over from the equilibrium Johnson-Nyquist value $4k_BT_0/R$ to a remarkably linear shot noise dependence $S_I=2eFI_{NW}$, that persists up to $|I_{NW}|\approx3\mu$A in longer devices or even up to $|I_{NW}|\approx10\mu$A in  shorter devices and is characterized by a Fano factor $F\approx0.3$, nearly the same for all NWs studied. Such a behavior is a hallmark of elastic diffusive transport, that is characterized by a universal value $F=1/3$~\cite{Nagaev1992,Beenakker_Buettiker_1992}. Note, that minor deviations of $F$ from 1/3 in our experiment can be caused by mesoscopic shot noise fluctuations \cite{deJong_Beenakker_1992} and/or a calibration uncertainty. Such mesoscopic fluctuations are clearly visible as slight irregularities of the slope for the short section data, see the circles in Fig.~\ref{fig2}. The elastic diffusive transport regime in our NWs was observed to break down only in longer devices for $|I_{NW}|\geq3\mu A$, corresponding to a noise temperature above $40$\,K. Here, a gradual deflection of $S_I$ from the linear dependence becomes evident at increasing $|I_{NW}|$, that results from the electron energy relaxation via  acoustic phonon emission and can be used to estimate the inelastic scattering length~\cite{Nagaev1992}. Regarding TE experiments, this indicates that a mutual impact of non-equilibrium phononic and electronic NW subsystems, including possible electron-phonon drag effects, can not be neglected for such high temperatures. Below we concentrate on a temperature range around $T_0=4.2$~K where such effects are not important and the TE experiment probes the diffusion thermopower.

{\bf Measurement of the NW thermal bias.}

It is convenient to treat diffusive transport and noise in our NWs within the quasi-classical approach~\cite{Nagaev1992} by means of local electronic energy distribution $f_{\varepsilon,x}$. For elastic diffusion the kinetic equation for $f_{\varepsilon,x}$ reduces to: 
\begin{equation}
\frac{\partial^2f_{\varepsilon,x}}{\partial x^2}=0,
\label{delta_f}\end{equation}
where $x$ is a coordinate along the NW and $\varepsilon$ is the electron energy relative to the Fermi level. The solution of the eq.~(\ref{delta_f}) is simply $f_{\varepsilon,x}=f_{\varepsilon,0}(1-x/L)+f_{\varepsilon,L}(x/L)$, where $f_{\varepsilon,0}$  and $f_{\varepsilon,L}$ are the boundary conditions on the two ends of the NW. This solution also determines the spectral density of spontaneous current fluctuations in the NW via:
\begin{equation}
S_I=\frac{4}{R}\int{\frac{dx}{L}}\int{f_{\varepsilon,x}(1-f_{\varepsilon,x})d\varepsilon},
\label{spectral_density}
\end{equation}
where $R$ is the NW resistance. The eqs.~(\ref{spectral_density}) and~(\ref{delta_f}) also define the NW noise temperature $T_{NW}$ via a Johnson-Nyquist like relation  $T_{NW}=\int{T_N(x)\frac{dx}{L}}$, where $T_N(x)=(k_B)^{-1}\int{f_{\varepsilon,x}(1-f_{\varepsilon,x})d\varepsilon}$ is the noise temperature for a given position. For a uniform Fermi-Dirac distribution $T_{NW}$ reduces to the usual equilibrium temperature. Below, we use the eqs.~(\ref{delta_f}) and~(\ref{spectral_density}) to quantify the thermal bias across an individual NW. 

In our TE experiments we mostly used a contact heating scheme, when the thermal bias $\delta T$ across the NW is created by means of the heating current $I_H$, e.g. flowing between the terminals 1 and 3 via the contact heater H1 in Fig.~\ref{fig1}a. Similar approach was used in a thermoelectric experiment of Ref.~\cite{Strunk1998}. A conceptual advantage of the present experiment, however, is that we use the noise thermometry to directly characterize the NW device under test, rather than to measure the average temperature of the metallic heater. The heating current modifies the energy distribution $f_{\varepsilon,0}=f_0+\delta f_H$ at the hot-end of the NW ($x=0$), whereas the opposite cold-end of the NW  ($x=L$) remains in equilibrium $f_{\varepsilon,L}=f_0\equiv(\exp{(\varepsilon/k_BT_0)}+1)^{-1}$. The modified distribution $f_{\varepsilon,0}$ is not necessary thermal~\cite{Pothier1997}, therefore we define the thermal bias via the excess noise temperature on the hot-end $\delta T\equiv T_N(0)-T_0$. For small enough $I_H$ a relation between $\delta T$ and the  noise temperature of the NW can be derived with the eqs.~(\ref{delta_f}) and~(\ref{spectral_density}), namely:
\begin{equation}
	\delta T_{NW}=\delta T/2,
	\label{dT}
\end{equation}
where we introduced excess noise temperature of the NW $\delta T_{NW}=T_{NW}-T_0$. Note that this intuitive relation
 holds for elastic diffusive transport and an arbitrary energy distribution on the hot-end provided $\delta f_H\ll f_0$. 

\begin{figure}[t]
\begin{center}
\vspace{10mm}
  \includegraphics[width=0.8\columnwidth]{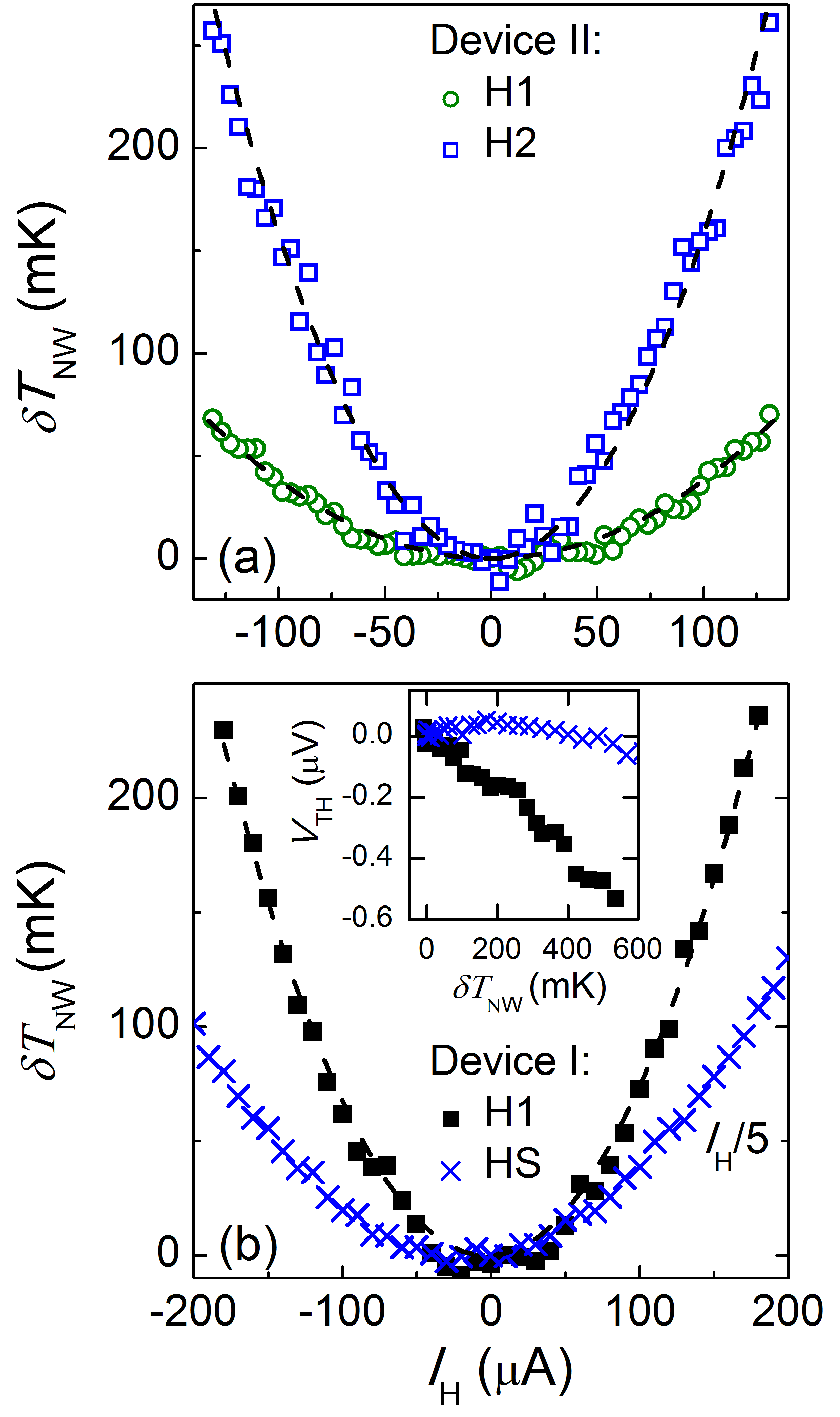}
	   \end{center}
  \caption{(a) Device~II: Noise temperature change $\delta T_{NW}$ of the short (circles) and long (squares) sections of the NW in response to the heating current~$I_H$ through the contact heaters H1 and H2, respectively. (b) Device~I: Noise temperature change $\delta T_{NW}$ of the NW in response to the heating current~$I_H$ through the contact stripe heater H1 (squares) and through the substrate heater HS (crosses), respectively. Dashed guide lines reflect the parabolic dependencies. Inset: Comparison of induced TE voltages for contact (squares) and substrate (crosses) heaters for the same range of $\delta T_{NW}$ values.}
	\label{fig3}	
\end{figure}

Eq.~(\ref{dT}) enables us to quantify the thermal bias across the NW by means of the noise thermometry~\cite{noise_sensor}. In fig.~\ref{fig3}a we plot the measured $\delta T_{NW}$ in dependence on $I_H$ for the two sections of the NW device II. This experiment allows to compare the heating efficiencies of the short and narrow contact heater H1 versus the long and wide contact heater H2 attached, respectively, to the short and long NW sections, see fig.~\ref{fig1}b. For both the short section (circles) and the long section (squares) parabolic dependencies are observed as demonstrated by the dashed line fits of the form $\delta T\propto (I_H)^2$. This illustrates the fact that for small $\delta T$ the temperature rise is proportional to the amount of the released Joule heat. Note, however, that in spite of a factor of $\sim10$ difference between the two heater resistances the corresponding $\delta T$ in Fig.~\ref{fig3}a differ only by a factor of $\sim4$. The reason is a smaller Joule heating efficiency of the long and wide heater H2 owing to the electron-phonon energy loss, that is proportional to a heater volume and appears to be negligible for a short and narrow heater H1, see Ref.~\cite{noise_sensor}. 

In fig.~\ref{fig3}b we compare the measured $\delta T_{NW}$ for different heater types. Squares and crosses correspond, respectively, to the contact heater H1 and the substrate heater HS of the NW device I depicted in fig.~\ref{fig1}a (both heaters have about 3\,$\Omega$ resistances). Again, the data follow a nearly parabolic functional dependence, as indicated by the dashed line for the contact heater.
While the efficiency of the contact heater is comparable to the data of fig.~\ref{fig3}a, the substrate heater is found much less efficient, so that we had to reduce the corresponding abscissa scale by a factor of 5. This emphasizes a relatively weak electron-phonon coupling in our devices, on which the substrate heating relies. In addition, we observe that unlike contact heating, the substrate heating is less effective in creating the thermal bias across the NW. This is demonstrated by TE measurements in the inset of fig.~\ref{fig3}b. Here, we plot the TE voltage $V_{th}$ across the NW, which is identified as the voltage drop between the terminals N and 2 in device I symmetrized for positive and negative $I_H$, see Fig.~\ref{fig1}. We find that for the same $\delta T_{NW}$, the $|V_{th}|$ is much smaller when the substrate heater is used instead of the contact heater (crosses vs squares, respectively). Thus, the substrate heater tends to heat up the NW as a whole, which is not the case for the contact heating configuration. For this reason, in the following we concentrate on the TE measurements using the contact heaters.

{\bf Thermoelectric measurements}

\begin{figure*}[t]
\begin{center}
\vspace{10mm}
  \includegraphics[width=0.9\linewidth]{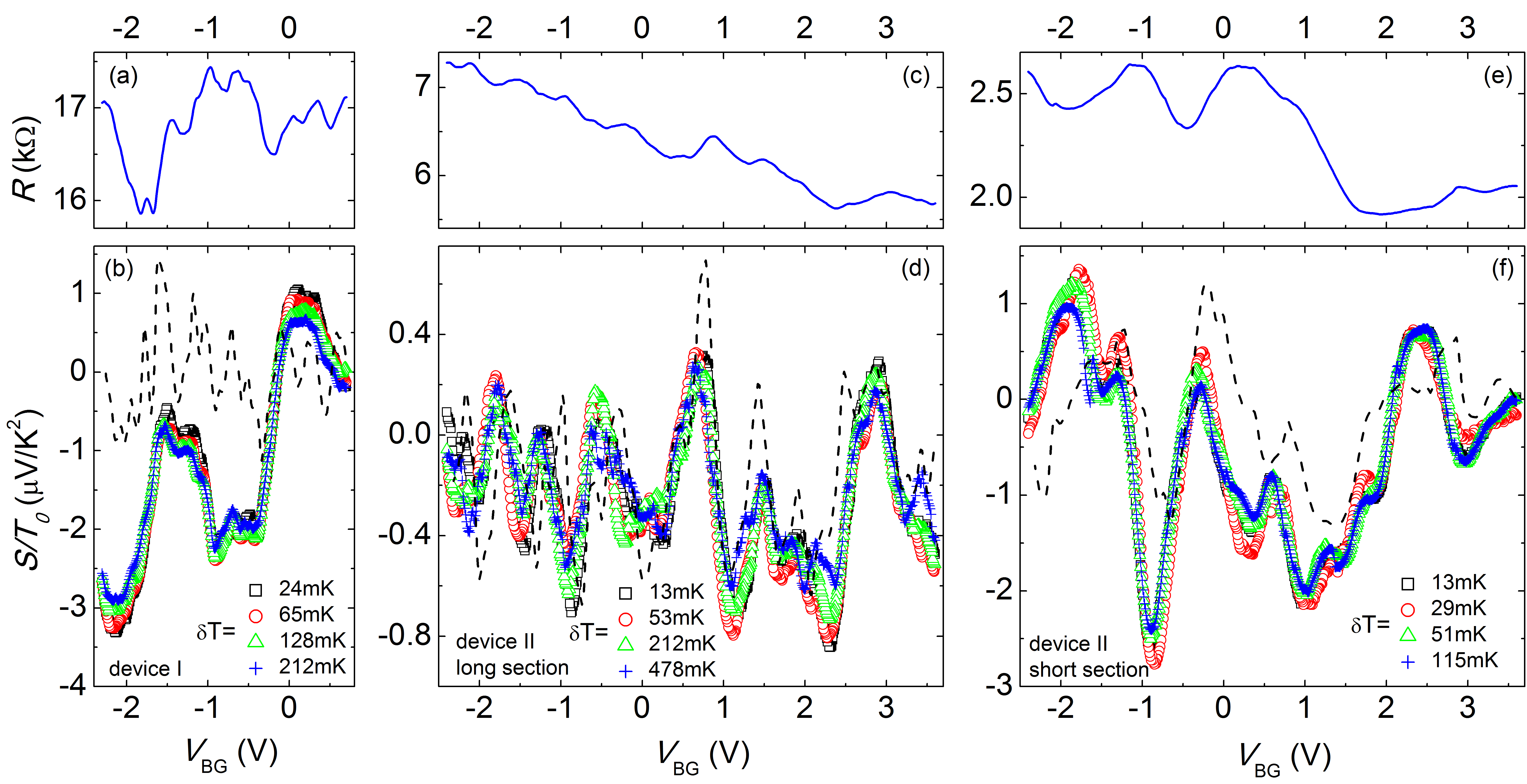}
\end{center}
  \caption{Linear response resistance and normalized Seebeck coefficient $S/T_0$ as a function of back-gate voltage measured on device~I (a,b) and device~II for short (c,d) and long (e,f) NW sections. Different symbols correspond to different values of the thermal bias, see respective legends. The dashed lines are numerically calculated from the data $R(V_{BG})$ and the Mott's thermopower relation using a gate-voltage to the Fermi energy conversion factor  $\alpha\sim\rm6.7\,meV/V$, see text.}
	\label{fig4}
\end{figure*}

The applied thermal bias results in a TE voltage drop $V_{th}$ between the (equilibrium) contact N and the heated end of the NW, which is kept open circuit during the TE measurement. Experimentally, $V_{th}$ is a tiny contribution masked by a resistive voltage drop across the part of the current biased contact stripe heater, which is involved in this measurement. For instance in the device I, the resistive contribution drops between the terminals 2 and N, see fig~\ref{fig1}a. In the following we modulate $I_H$ with a small ac current 20-70~nA at a frequency of 11\,Hz, that corresponds to $\delta T$ in the range 10-500\,mK, and measure the induced $V_{th}$ using a lock-in second harmonic detection. We convert the data into the Seebeck coefficient $S\equiv V_{th}/\delta T$, normalize it by the bath temperature and plot $S/T_0$. Figs.~\ref{fig4}b,d,f demonstrate, respectively, the data for the device I, and for long and short NW sections of the device II in dependence of the back-gate voltage $V_{BG}$. In all three cases we observe the same qualitative behavior. Within a broad range of $\delta T$ the datasets in each panel collapse on a single curve, justifying the linear dependence of $V_{th}$ on $\delta T$.  On the average, the TE signal is apparently negative, on the order of $\overline{S}/T_0\sim-1\div-0.2\,\mu{\rm V/K^2}$, as expected for n-type charge carriers in our InAs NWs and consistent with previous studies~\cite{Roddaro2013}. Yet, the overall TE signal is dominated by pronounced mesoscopic fluctuations that even cause a sign reversal of $S$ in certain gate-voltage ranges. The mesoscopic origin of the fluctuations is consistent with the two following observations. First, the strongest fluctuations are found in the most resistive device I. Second, the fluctuations are weaker in the longer section of the device II, consistently with the length dependent self-averaging. Note also, that our NWs are highly doped and characterized by a relatively small resistance and  diffusive  transport mechanism, as verified via shot noise, see Fig.~\ref{fig2}. This is in contrast to the lower doping NWs for which the fluctuations of $S$ were interpreted in terms of quantum dot like states~\cite{Wu2013}.

Below we compare the TE measurements with the behavior of the gate-voltage dependent resistance $R$. As shown in Figs.~\ref{fig4}a,c and e for the respective measurement configurations, the measured $R$ exhibits small random fluctuations as a function of $V_{BG}$, that tunes the carrier density and the chemical potential of the NW electron system. According to the Mott's thermopower law~\cite{Mott_Jones} the energy dependence of the conductivity and the TE response are related as $S/T_0=-(\pi k_B)^2/(3e\sigma)d\sigma/dE_F$, where the derivative is evaluated at the Fermi energy. Assuming a linear dependence of $E_F$ on $V_{BG}$, in Figs.~\ref{fig4}b,d and f we plot the numerically calculated Mott's law $(\pi k_B)^2/(3eR\alpha) dR/dV_{BG}$ (dashed lines), where $\alpha=dE_F/dV_{BG}$. The results differ for the two devices. For the device II, both short and long sections, the fluctuations of $S$ are correlated with the Mott's law data, see Figs.~\ref{fig4}d and f. This similarity enables us to evaluate a gate-voltage to Fermi energy conversion factor at $\alpha\sim\rm7\,meV/V$. This value of $\alpha$ is consistent with estimates of the density of states and the back gate capacitance in our NWs, as well as with the assumption of slow variation of $\sigma(E)$ on the scale of $k_BT_0$, which is implied by the Mott's law. By contrast, no obvious correlation between the measured and evaluated $S$ is observed for the device I (Fig.~\ref{fig4}b). Moreover, the typical $V_{BG}$ scale of the resistance fluctuations is considerably shorter than the one for the TE signal. Most probably, this is a consequence of a much stronger impact of charge carrier traps on the gate-voltage swept resistance data in this sample, which caused hysteresis and was the reason for the narrower sweep range in the device I. 

{\bf Summary}

In summary, we applied a primary noise thermometry to investigate charge and TE transport in individual InAs NWs. This served to identify elastic diffusive transport regime in our devices with negligible electron-phonon interaction at low temperatures. In TE measurements, the noise thermometry enabled us to use a contact heating approach, that turned out much more efficient than conventional substrate heating in creating a thermal bias across the NWs. With this approach, we measured the Seebeck coefficient $S$ in two devices at $T_0=4.2$~K in dependence on the back-gate voltage. We observed pronounced random mesoscopic fluctuations of $S$, identified their rough correlation with the mesoscopic resistance fluctuations via the Mott's thermopower law in one device and evaluate a gate-voltage to Fermi energy conversion factor. Our results demonstrate the primary noise thermometry as a powerful tool for mesoscopic thermal transport applications, which is perfectly compatible with standard measurement techniques.

	
\bibliography{InAsbbl}

\begin{thebibliography}{23}%
\makeatletter
\providecommand \@ifxundefined [1]{%
 \@ifx{#1\undefined}
}%
\providecommand \@ifnum [1]{%
 \ifnum #1\expandafter \@firstoftwo
 \else \expandafter \@secondoftwo
 \fi
}%
\providecommand \@ifx [1]{%
 \ifx #1\expandafter \@firstoftwo
 \else \expandafter \@secondoftwo
 \fi
}%
\providecommand \natexlab [1]{#1}%
\providecommand \enquote  [1]{``#1''}%
\providecommand \bibnamefont  [1]{#1}%
\providecommand \bibfnamefont [1]{#1}%
\providecommand \citenamefont [1]{#1}%
\providecommand \href@noop [0]{\@secondoftwo}%
\providecommand \href [0]{\begingroup \@sanitize@url \@href}%
\providecommand \@href[1]{\@@startlink{#1}\@@href}%
\providecommand \@@href[1]{\endgroup#1\@@endlink}%
\providecommand \@sanitize@url [0]{\catcode `\\12\catcode `\$12\catcode
  `\&12\catcode `\#12\catcode `\^12\catcode `\_12\catcode `\%12\relax}%
\providecommand \@@startlink[1]{}%
\providecommand \@@endlink[0]{}%
\providecommand \url  [0]{\begingroup\@sanitize@url \@url }%
\providecommand \@url [1]{\endgroup\@href {#1}{\urlprefix }}%
\providecommand \urlprefix  [0]{URL }%
\providecommand \Eprint [0]{\href }%
\providecommand \doibase [0]{http://dx.doi.org/}%
\providecommand \selectlanguage [0]{\@gobble}%
\providecommand \bibinfo  [0]{\@secondoftwo}%
\providecommand \bibfield  [0]{\@secondoftwo}%
\providecommand \translation [1]{[#1]}%
\providecommand \BibitemOpen [0]{}%
\providecommand \bibitemStop [0]{}%
\providecommand \bibitemNoStop [0]{.\EOS\space}%
\providecommand \EOS [0]{\spacefactor3000\relax}%
\providecommand \BibitemShut  [1]{\csname bibitem#1\endcsname}%
\let\auto@bib@innerbib\@empty
\bibitem [{\citenamefont {Snyder}\ and\ \citenamefont
  {Toberer}(2008)}]{Snyder2008}%
  \BibitemOpen
  \bibfield  {author} {\bibinfo {author} {\bibfnamefont {G.~J.}\ \bibnamefont
  {Snyder}}\ and\ \bibinfo {author} {\bibfnamefont {E.~S.}\ \bibnamefont
  {Toberer}},\ }\href {\doibase 10.1038/nmat2090} {\bibfield  {journal}
  {\bibinfo  {journal} {Nature Materials}\ }\textbf {\bibinfo {volume} {7}},\
  \bibinfo {pages} {105} (\bibinfo {year} {2008})}\BibitemShut {NoStop}%
\bibitem [{\citenamefont {Majumdar}(2004)}]{Majumdar2004}%
  \BibitemOpen
  \bibfield  {author} {\bibinfo {author} {\bibfnamefont {A.}~\bibnamefont
  {Majumdar}},\ }\href {\doibase 10.1126/science.1093164} {\bibfield  {journal}
  {\bibinfo  {journal} {Science}\ }\textbf {\bibinfo {volume} {303}},\ \bibinfo
  {pages} {777} (\bibinfo {year} {2004})}\BibitemShut {NoStop}%
\bibitem [{\citenamefont {Dresselhaus}\ \emph {et~al.}(2007)\citenamefont
  {Dresselhaus}, \citenamefont {Chen}, \citenamefont {Tang}, \citenamefont
  {Yang}, \citenamefont {Lee}, \citenamefont {Wang}, \citenamefont {Ren},
  \citenamefont {Fleurial},\ and\ \citenamefont {Gogna}}]{Dresselhaus2007}%
  \BibitemOpen
  \bibfield  {author} {\bibinfo {author} {\bibfnamefont {M.~S.}\ \bibnamefont
  {Dresselhaus}}, \bibinfo {author} {\bibfnamefont {G.}~\bibnamefont {Chen}},
  \bibinfo {author} {\bibfnamefont {M.~Y.}\ \bibnamefont {Tang}}, \bibinfo
  {author} {\bibfnamefont {R.~G.}\ \bibnamefont {Yang}}, \bibinfo {author}
  {\bibfnamefont {H.}~\bibnamefont {Lee}}, \bibinfo {author} {\bibfnamefont
  {D.~Z.}\ \bibnamefont {Wang}}, \bibinfo {author} {\bibfnamefont {Z.~F.}\
  \bibnamefont {Ren}}, \bibinfo {author} {\bibfnamefont {J.-P.}\ \bibnamefont
  {Fleurial}}, \ and\ \bibinfo {author} {\bibfnamefont {P.}~\bibnamefont
  {Gogna}},\ }\href {\doibase 10.1002/adma.200600527} {\bibfield  {journal}
  {\bibinfo  {journal} {Adv. Mater.}\ }\textbf {\bibinfo {volume} {19}},\
  \bibinfo {pages} {1043} (\bibinfo {year} {2007})}\BibitemShut {NoStop}%
\bibitem [{\citenamefont {Shi}(2012)}]{Shi2012}%
  \BibitemOpen
  \bibfield  {author} {\bibinfo {author} {\bibfnamefont {L.}~\bibnamefont
  {Shi}},\ }\href {\doibase 10.1080/15567265.2012.667514} {\bibfield  {journal}
  {\bibinfo  {journal} {Nanoscale and Microscale Thermophysical Engineering}\
  }\textbf {\bibinfo {volume} {16}},\ \bibinfo {pages} {79} (\bibinfo {year}
  {2012})}\BibitemShut {NoStop}%
\bibitem [{\citenamefont {Boukai}\ \emph {et~al.}(2008)\citenamefont {Boukai},
  \citenamefont {Bunimovich}, \citenamefont {Tahir-Kheli}, \citenamefont {Yu},
  \citenamefont {III},\ and\ \citenamefont {Heath}}]{Boukai2008}%
  \BibitemOpen
  \bibfield  {author} {\bibinfo {author} {\bibfnamefont {A.~I.}\ \bibnamefont
  {Boukai}}, \bibinfo {author} {\bibfnamefont {Y.}~\bibnamefont {Bunimovich}},
  \bibinfo {author} {\bibfnamefont {J.}~\bibnamefont {Tahir-Kheli}}, \bibinfo
  {author} {\bibfnamefont {J.-K.}\ \bibnamefont {Yu}}, \bibinfo {author}
  {\bibfnamefont {W.~A.~G.}\ \bibnamefont {III}}, \ and\ \bibinfo {author}
  {\bibfnamefont {J.~R.}\ \bibnamefont {Heath}},\ }\href {\doibase
  10.1038/nature06458} {\bibfield  {journal} {\bibinfo  {journal} {Nature}\
  }\textbf {\bibinfo {volume} {451}},\ \bibinfo {pages} {168} (\bibinfo {year}
  {2008})}\BibitemShut {NoStop}%
\bibitem [{\citenamefont {Hochbaum}\ \emph {et~al.}(2008)\citenamefont
  {Hochbaum}, \citenamefont {Chen}, \citenamefont {Delgado}, \citenamefont
  {Liang}, \citenamefont {Garnett}, \citenamefont {Najarian}, \citenamefont
  {Majumdar},\ and\ \citenamefont {Yang}}]{Hochbaum2008}%
  \BibitemOpen
  \bibfield  {author} {\bibinfo {author} {\bibfnamefont {A.~I.}\ \bibnamefont
  {Hochbaum}}, \bibinfo {author} {\bibfnamefont {R.}~\bibnamefont {Chen}},
  \bibinfo {author} {\bibfnamefont {R.~D.}\ \bibnamefont {Delgado}}, \bibinfo
  {author} {\bibfnamefont {W.}~\bibnamefont {Liang}}, \bibinfo {author}
  {\bibfnamefont {E.~C.}\ \bibnamefont {Garnett}}, \bibinfo {author}
  {\bibfnamefont {M.}~\bibnamefont {Najarian}}, \bibinfo {author}
  {\bibfnamefont {A.}~\bibnamefont {Majumdar}}, \ and\ \bibinfo {author}
  {\bibfnamefont {P.}~\bibnamefont {Yang}},\ }\href {\doibase
  10.1038/nature06381} {\bibfield  {journal} {\bibinfo  {journal} {Nature}\
  }\textbf {\bibinfo {volume} {451}},\ \bibinfo {pages} {163} (\bibinfo {year}
  {2008})}\BibitemShut {NoStop}%
\bibitem [{\citenamefont {Martinez}\ \emph {et~al.}(2011)\citenamefont
  {Martinez}, \citenamefont {Provencio}, \citenamefont {Picraux}, \citenamefont
  {Sullivan},\ and\ \citenamefont {Swartzentruber}}]{Martinez2011}%
  \BibitemOpen
  \bibfield  {author} {\bibinfo {author} {\bibfnamefont {J.~A.}\ \bibnamefont
  {Martinez}}, \bibinfo {author} {\bibfnamefont {P.~P.}\ \bibnamefont
  {Provencio}}, \bibinfo {author} {\bibfnamefont {S.~T.}\ \bibnamefont
  {Picraux}}, \bibinfo {author} {\bibfnamefont {J.~P.}\ \bibnamefont
  {Sullivan}}, \ and\ \bibinfo {author} {\bibfnamefont {B.~S.}\ \bibnamefont
  {Swartzentruber}},\ }\href {\doibase 10.1063/1.3647575} {\bibfield  {journal}
  {\bibinfo  {journal} {J. Appl. Phys.}\ }\textbf {\bibinfo {volume} {110}},\
  \bibinfo {pages} {074317} (\bibinfo {year} {2011})}\BibitemShut {NoStop}%
\bibitem [{\citenamefont {Swinkels}\ \emph {et~al.}(2015)\citenamefont
  {Swinkels}, \citenamefont {van Delft}, \citenamefont {Oliveira},
  \citenamefont {Cavalli}, \citenamefont {Zardo}, \citenamefont {van~der
  Heijden},\ and\ \citenamefont {Bakkers}}]{Swinkels2015}%
  \BibitemOpen
  \bibfield  {author} {\bibinfo {author} {\bibfnamefont {M.~Y.}\ \bibnamefont
  {Swinkels}}, \bibinfo {author} {\bibfnamefont {M.~R.}\ \bibnamefont {van
  Delft}}, \bibinfo {author} {\bibfnamefont {D.~S.}\ \bibnamefont {Oliveira}},
  \bibinfo {author} {\bibfnamefont {A.}~\bibnamefont {Cavalli}}, \bibinfo
  {author} {\bibfnamefont {I.}~\bibnamefont {Zardo}}, \bibinfo {author}
  {\bibfnamefont {R.~W.}\ \bibnamefont {van~der Heijden}}, \ and\ \bibinfo
  {author} {\bibfnamefont {E.~P. A.~M.}\ \bibnamefont {Bakkers}},\ }\href
  {\doibase 10.1088/0957-4484/26/38/385401} {\bibfield  {journal} {\bibinfo
  {journal} {Nanotechnology}\ }\textbf {\bibinfo {volume} {26}},\ \bibinfo
  {pages} {385401} (\bibinfo {year} {2015})}\BibitemShut {NoStop}%
\bibitem [{\citenamefont {Zhou}\ \emph {et~al.}(2011)\citenamefont {Zhou},
  \citenamefont {Moore}, \citenamefont {Bolinsson}, \citenamefont {Persson},
  \citenamefont {Fr\"{o}berg}, \citenamefont {Pettes}, \citenamefont {Kong},
  \citenamefont {Rabenberg}, \citenamefont {Caroff}, \citenamefont {Stewart},
  \citenamefont {Mingo}, \citenamefont {Dick}, \citenamefont {Samuelson},
  \citenamefont {Linke},\ and\ \citenamefont {Shi}}]{Zhou2011}%
  \BibitemOpen
  \bibfield  {author} {\bibinfo {author} {\bibfnamefont {F.}~\bibnamefont
  {Zhou}}, \bibinfo {author} {\bibfnamefont {A.~L.}\ \bibnamefont {Moore}},
  \bibinfo {author} {\bibfnamefont {J.}~\bibnamefont {Bolinsson}}, \bibinfo
  {author} {\bibfnamefont {A.}~\bibnamefont {Persson}}, \bibinfo {author}
  {\bibfnamefont {L.}~\bibnamefont {Fr\"{o}berg}}, \bibinfo {author}
  {\bibfnamefont {M.~T.}\ \bibnamefont {Pettes}}, \bibinfo {author}
  {\bibfnamefont {H.}~\bibnamefont {Kong}}, \bibinfo {author} {\bibfnamefont
  {L.}~\bibnamefont {Rabenberg}}, \bibinfo {author} {\bibfnamefont
  {P.}~\bibnamefont {Caroff}}, \bibinfo {author} {\bibfnamefont {D.~A.}\
  \bibnamefont {Stewart}}, \bibinfo {author} {\bibfnamefont {N.}~\bibnamefont
  {Mingo}}, \bibinfo {author} {\bibfnamefont {K.~A.}\ \bibnamefont {Dick}},
  \bibinfo {author} {\bibfnamefont {L.}~\bibnamefont {Samuelson}}, \bibinfo
  {author} {\bibfnamefont {H.}~\bibnamefont {Linke}}, \ and\ \bibinfo {author}
  {\bibfnamefont {L.}~\bibnamefont {Shi}},\ }\href {\doibase
  10.1103/physrevb.83.205416} {\bibfield  {journal} {\bibinfo  {journal} {Phys.
  Rev. B}\ }\textbf {\bibinfo {volume} {83}} (\bibinfo {year} {2011}),\
  10.1103/physrevb.83.205416}\BibitemShut {NoStop}%
\bibitem [{\citenamefont {Heremans}\ \emph {et~al.}(2008)\citenamefont
  {Heremans}, \citenamefont {Jovovic}, \citenamefont {Toberer}, \citenamefont
  {Saramat}, \citenamefont {Kurosaki}, \citenamefont {Charoenphakdee},
  \citenamefont {Yamanaka},\ and\ \citenamefont {Snyder}}]{Heremans2008}%
  \BibitemOpen
  \bibfield  {author} {\bibinfo {author} {\bibfnamefont {J.~P.}\ \bibnamefont
  {Heremans}}, \bibinfo {author} {\bibfnamefont {V.}~\bibnamefont {Jovovic}},
  \bibinfo {author} {\bibfnamefont {E.~S.}\ \bibnamefont {Toberer}}, \bibinfo
  {author} {\bibfnamefont {A.}~\bibnamefont {Saramat}}, \bibinfo {author}
  {\bibfnamefont {K.}~\bibnamefont {Kurosaki}}, \bibinfo {author}
  {\bibfnamefont {A.}~\bibnamefont {Charoenphakdee}}, \bibinfo {author}
  {\bibfnamefont {S.}~\bibnamefont {Yamanaka}}, \ and\ \bibinfo {author}
  {\bibfnamefont {G.~J.}\ \bibnamefont {Snyder}},\ }\href {\doibase
  10.1126/science.1159725} {\bibfield  {journal} {\bibinfo  {journal}
  {Science}\ }\textbf {\bibinfo {volume} {321}},\ \bibinfo {pages} {554}
  (\bibinfo {year} {2008})}\BibitemShut {NoStop}%
\bibitem [{\citenamefont {Tian}\ \emph {et~al.}(2012)\citenamefont {Tian},
  \citenamefont {Sakr}, \citenamefont {Kinder}, \citenamefont {Liang},
  \citenamefont {MacDonald}, \citenamefont {Qiu}, \citenamefont {Gao},\ and\
  \citenamefont {Gao}}]{Tian2012}%
  \BibitemOpen
  \bibfield  {author} {\bibinfo {author} {\bibfnamefont {Y.}~\bibnamefont
  {Tian}}, \bibinfo {author} {\bibfnamefont {M.~R.}\ \bibnamefont {Sakr}},
  \bibinfo {author} {\bibfnamefont {J.~M.}\ \bibnamefont {Kinder}}, \bibinfo
  {author} {\bibfnamefont {D.}~\bibnamefont {Liang}}, \bibinfo {author}
  {\bibfnamefont {M.~J.}\ \bibnamefont {MacDonald}}, \bibinfo {author}
  {\bibfnamefont {R.~L.~J.}\ \bibnamefont {Qiu}}, \bibinfo {author}
  {\bibfnamefont {H.-J.}\ \bibnamefont {Gao}}, \ and\ \bibinfo {author}
  {\bibfnamefont {X.~P.~A.}\ \bibnamefont {Gao}},\ }\href {\doibase
  10.1021/nl304194c} {\bibfield  {journal} {\bibinfo  {journal} {Nano Letters}\
  }\textbf {\bibinfo {volume} {12}},\ \bibinfo {pages} {6492} (\bibinfo {year}
  {2012})}\BibitemShut {NoStop}%
\bibitem [{\citenamefont {Wu}\ \emph {et~al.}(2013)\citenamefont {Wu},
  \citenamefont {Gooth}, \citenamefont {Zianni}, \citenamefont {Svensson},
  \citenamefont {Gluschke}, \citenamefont {Dick}, \citenamefont {Thelander},
  \citenamefont {Nielsch},\ and\ \citenamefont {Linke}}]{Wu2013}%
  \BibitemOpen
  \bibfield  {author} {\bibinfo {author} {\bibfnamefont {P.~M.}\ \bibnamefont
  {Wu}}, \bibinfo {author} {\bibfnamefont {J.}~\bibnamefont {Gooth}}, \bibinfo
  {author} {\bibfnamefont {X.}~\bibnamefont {Zianni}}, \bibinfo {author}
  {\bibfnamefont {S.~F.}\ \bibnamefont {Svensson}}, \bibinfo {author}
  {\bibfnamefont {J.~G.}\ \bibnamefont {Gluschke}}, \bibinfo {author}
  {\bibfnamefont {K.~A.}\ \bibnamefont {Dick}}, \bibinfo {author}
  {\bibfnamefont {C.}~\bibnamefont {Thelander}}, \bibinfo {author}
  {\bibfnamefont {K.}~\bibnamefont {Nielsch}}, \ and\ \bibinfo {author}
  {\bibfnamefont {H.}~\bibnamefont {Linke}},\ }\href {\doibase
  10.1021/nl401501j} {\bibfield  {journal} {\bibinfo  {journal} {Nano Letters}\
  }\textbf {\bibinfo {volume} {13}},\ \bibinfo {pages} {4080} (\bibinfo {year}
  {2013})}\BibitemShut {NoStop}%
\bibitem [{\citenamefont {Hoffmann}\ \emph {et~al.}(2009)\citenamefont
  {Hoffmann}, \citenamefont {Nilsson}, \citenamefont {Matthews}, \citenamefont
  {Nakpathomkun}, \citenamefont {Persson}, \citenamefont {Samuelson},\ and\
  \citenamefont {Linke}}]{Hoffmann2009}%
  \BibitemOpen
  \bibfield  {author} {\bibinfo {author} {\bibfnamefont {E.~A.}\ \bibnamefont
  {Hoffmann}}, \bibinfo {author} {\bibfnamefont {H.~A.}\ \bibnamefont
  {Nilsson}}, \bibinfo {author} {\bibfnamefont {J.~E.}\ \bibnamefont
  {Matthews}}, \bibinfo {author} {\bibfnamefont {N.}~\bibnamefont
  {Nakpathomkun}}, \bibinfo {author} {\bibfnamefont {A.~I.}\ \bibnamefont
  {Persson}}, \bibinfo {author} {\bibfnamefont {L.}~\bibnamefont {Samuelson}},
  \ and\ \bibinfo {author} {\bibfnamefont {H.}~\bibnamefont {Linke}},\ }\href
  {\doibase 10.1021/nl8034042} {\bibfield  {journal} {\bibinfo  {journal} {Nano
  Letters}\ }\textbf {\bibinfo {volume} {9}},\ \bibinfo {pages} {779} (\bibinfo
  {year} {2009})}\BibitemShut {NoStop}%
\bibitem [{\citenamefont {Roddaro}\ \emph {et~al.}(2013)\citenamefont
  {Roddaro}, \citenamefont {Ercolani}, \citenamefont {Safeen}, \citenamefont
  {Suomalainen}, \citenamefont {Rossella}, \citenamefont {Giazotto},
  \citenamefont {Sorba},\ and\ \citenamefont {Beltram}}]{Roddaro2013}%
  \BibitemOpen
  \bibfield  {author} {\bibinfo {author} {\bibfnamefont {S.}~\bibnamefont
  {Roddaro}}, \bibinfo {author} {\bibfnamefont {D.}~\bibnamefont {Ercolani}},
  \bibinfo {author} {\bibfnamefont {M.~A.}\ \bibnamefont {Safeen}}, \bibinfo
  {author} {\bibfnamefont {S.}~\bibnamefont {Suomalainen}}, \bibinfo {author}
  {\bibfnamefont {F.}~\bibnamefont {Rossella}}, \bibinfo {author}
  {\bibfnamefont {F.}~\bibnamefont {Giazotto}}, \bibinfo {author}
  {\bibfnamefont {L.}~\bibnamefont {Sorba}}, \ and\ \bibinfo {author}
  {\bibfnamefont {F.}~\bibnamefont {Beltram}},\ }\href {\doibase
  10.1021/nl401482p} {\bibfield  {journal} {\bibinfo  {journal} {Nano Letters}\
  }\textbf {\bibinfo {volume} {13}},\ \bibinfo {pages} {3638} (\bibinfo {year}
  {2013})}\BibitemShut {NoStop}%
\bibitem [{\citenamefont {Yazji}\ \emph {et~al.}(2015)\citenamefont {Yazji},
  \citenamefont {Hoffman}, \citenamefont {Ercolani}, \citenamefont {Rossella},
  \citenamefont {Pitanti}, \citenamefont {Cavalli}, \citenamefont {Roddaro},
  \citenamefont {Abstreiter}, \citenamefont {Sorba},\ and\ \citenamefont
  {Zardo}}]{Yazji2015}%
  \BibitemOpen
  \bibfield  {author} {\bibinfo {author} {\bibfnamefont {S.}~\bibnamefont
  {Yazji}}, \bibinfo {author} {\bibfnamefont {E.~A.}\ \bibnamefont {Hoffman}},
  \bibinfo {author} {\bibfnamefont {D.}~\bibnamefont {Ercolani}}, \bibinfo
  {author} {\bibfnamefont {F.}~\bibnamefont {Rossella}}, \bibinfo {author}
  {\bibfnamefont {A.}~\bibnamefont {Pitanti}}, \bibinfo {author} {\bibfnamefont
  {A.}~\bibnamefont {Cavalli}}, \bibinfo {author} {\bibfnamefont
  {S.}~\bibnamefont {Roddaro}}, \bibinfo {author} {\bibfnamefont
  {G.}~\bibnamefont {Abstreiter}}, \bibinfo {author} {\bibfnamefont
  {L.}~\bibnamefont {Sorba}}, \ and\ \bibinfo {author} {\bibfnamefont
  {I.}~\bibnamefont {Zardo}},\ }\href {\doibase 10.1007/s12274-015-0906-8}
  {\bibfield  {journal} {\bibinfo  {journal} {Nano Res.}\ }\textbf {\bibinfo
  {volume} {8}},\ \bibinfo {pages} {4048} (\bibinfo {year} {2015})}\BibitemShut
  {NoStop}%
\bibitem [{\citenamefont {Giazotto}\ \emph {et~al.}(2006)\citenamefont
  {Giazotto}, \citenamefont {Heikkil\"a}, \citenamefont {Luukanen},
  \citenamefont {Savin},\ and\ \citenamefont {Pekola}}]{Giazotto_Pekola_2006}%
  \BibitemOpen
  \bibfield  {author} {\bibinfo {author} {\bibfnamefont {F.}~\bibnamefont
  {Giazotto}}, \bibinfo {author} {\bibfnamefont {T.~T.}\ \bibnamefont
  {Heikkil\"a}}, \bibinfo {author} {\bibfnamefont {A.}~\bibnamefont
  {Luukanen}}, \bibinfo {author} {\bibfnamefont {A.~M.}\ \bibnamefont {Savin}},
  \ and\ \bibinfo {author} {\bibfnamefont {J.~P.}\ \bibnamefont {Pekola}},\
  }\href {\doibase 10.1103/RevModPhys.78.217} {\bibfield  {journal} {\bibinfo
  {journal} {Rev. Mod. Phys.}\ }\textbf {\bibinfo {volume} {78}},\ \bibinfo
  {pages} {217} (\bibinfo {year} {2006})}\BibitemShut {NoStop}%
\bibitem [{\citenamefont {Nagaev}(1992)}]{Nagaev1992}%
  \BibitemOpen
  \bibfield  {author} {\bibinfo {author} {\bibfnamefont {K.~E.}\ \bibnamefont
  {Nagaev}},\ }\href {\doibase http://dx.doi.org/10.1016/0375-9601(92)90814-3}
  {\bibfield  {journal} {\bibinfo  {journal} {Physics Letters A}\ }\textbf
  {\bibinfo {volume} {169}},\ \bibinfo {pages} {103 } (\bibinfo {year}
  {1992})}\BibitemShut {NoStop}%
\bibitem [{\citenamefont {Beenakker}\ and\ \citenamefont
  {B\"uttiker}(1992)}]{Beenakker_Buettiker_1992}%
  \BibitemOpen
  \bibfield  {author} {\bibinfo {author} {\bibfnamefont {C.~W.~J.}\
  \bibnamefont {Beenakker}}\ and\ \bibinfo {author} {\bibfnamefont
  {M.}~\bibnamefont {B\"uttiker}},\ }\href {\doibase 10.1103/PhysRevB.46.1889}
  {\bibfield  {journal} {\bibinfo  {journal} {Phys. Rev. B}\ }\textbf {\bibinfo
  {volume} {46}},\ \bibinfo {pages} {1889} (\bibinfo {year}
  {1992})}\BibitemShut {NoStop}%
\bibitem [{\citenamefont {de~Jong}\ and\ \citenamefont
  {Beenakker}(1992)}]{deJong_Beenakker_1992}%
  \BibitemOpen
  \bibfield  {author} {\bibinfo {author} {\bibfnamefont {M.~J.~M.}\
  \bibnamefont {de~Jong}}\ and\ \bibinfo {author} {\bibfnamefont {C.~W.~J.}\
  \bibnamefont {Beenakker}},\ }\href {\doibase 10.1103/PhysRevB.46.13400}
  {\bibfield  {journal} {\bibinfo  {journal} {Phys. Rev. B}\ }\textbf {\bibinfo
  {volume} {46}},\ \bibinfo {pages} {13400} (\bibinfo {year}
  {1992})}\BibitemShut {NoStop}%
\bibitem [{\citenamefont {Strunk}\ \emph {et~al.}(1998)\citenamefont {Strunk},
  \citenamefont {Henny}, \citenamefont {Sch\"onenberger}, \citenamefont
  {Neuttiens},\ and\ \citenamefont {Van~Haesendonck}}]{Strunk1998}%
  \BibitemOpen
  \bibfield  {author} {\bibinfo {author} {\bibfnamefont {C.}~\bibnamefont
  {Strunk}}, \bibinfo {author} {\bibfnamefont {M.}~\bibnamefont {Henny}},
  \bibinfo {author} {\bibfnamefont {C.}~\bibnamefont {Sch\"onenberger}},
  \bibinfo {author} {\bibfnamefont {G.}~\bibnamefont {Neuttiens}}, \ and\
  \bibinfo {author} {\bibfnamefont {C.}~\bibnamefont {Van~Haesendonck}},\
  }\href {\doibase 10.1103/PhysRevLett.81.2982} {\bibfield  {journal} {\bibinfo
   {journal} {Phys. Rev. Lett.}\ }\textbf {\bibinfo {volume} {81}},\ \bibinfo
  {pages} {2982} (\bibinfo {year} {1998})}\BibitemShut {NoStop}%
\bibitem [{\citenamefont {Pothier}\ \emph {et~al.}(1997)\citenamefont
  {Pothier}, \citenamefont {Gu{\'{e}}ron}, \citenamefont {Birge}, \citenamefont
  {Esteve},\ and\ \citenamefont {Devoret}}]{Pothier1997}%
  \BibitemOpen
  \bibfield  {author} {\bibinfo {author} {\bibfnamefont {H.}~\bibnamefont
  {Pothier}}, \bibinfo {author} {\bibfnamefont {S.}~\bibnamefont
  {Gu{\'{e}}ron}}, \bibinfo {author} {\bibfnamefont {N.~O.}\ \bibnamefont
  {Birge}}, \bibinfo {author} {\bibfnamefont {D.}~\bibnamefont {Esteve}}, \
  and\ \bibinfo {author} {\bibfnamefont {M.~H.}\ \bibnamefont {Devoret}},\
  }\href {\doibase 10.1103/physrevlett.79.3490} {\bibfield  {journal} {\bibinfo
   {journal} {Phys. Rev. Lett.}\ }\textbf {\bibinfo {volume} {79}},\ \bibinfo
  {pages} {3490} (\bibinfo {year} {1997})}\BibitemShut {NoStop}%
\bibitem [{\citenamefont {Tikhonov}\ \emph {et~al.}(2016)\citenamefont
  {Tikhonov}, \citenamefont {Shovkun}, \citenamefont {Khrapai}, \citenamefont
  {Ercolani}, \citenamefont {Sorba},\ and\ \citenamefont
  {Roddaro}}]{noise_sensor}%
  \BibitemOpen
  \bibfield  {author} {\bibinfo {author} {\bibfnamefont {E.~S.}\ \bibnamefont
  {Tikhonov}}, \bibinfo {author} {\bibfnamefont {D.}~\bibnamefont {Shovkun}},
  \bibinfo {author} {\bibfnamefont {V.}~\bibnamefont {Khrapai}}, \bibinfo
  {author} {\bibfnamefont {D.}~\bibnamefont {Ercolani}}, \bibinfo {author}
  {\bibfnamefont {L.}~\bibnamefont {Sorba}}, \ and\ \bibinfo {author}
  {\bibfnamefont {S.}~\bibnamefont {Roddaro}},\ }\href@noop {} {\bibfield
  {journal} {\bibinfo  {journal} {preprint}\ } (\bibinfo {year}
  {2016})}\BibitemShut {NoStop}%
\bibitem [{\citenamefont {Mott}\ and\ \citenamefont
  {Jones}(1958)}]{Mott_Jones}%
  \BibitemOpen
  \bibfield  {author} {\bibinfo {author} {\bibfnamefont {N.~F.}\ \bibnamefont
  {Mott}}\ and\ \bibinfo {author} {\bibfnamefont {H.}~\bibnamefont {Jones}},\
  }\href
  {http://www.amazon.com/The-Theory-Properties-Metals-Alloys/dp/048660456X%3FSubscriptionId%3D0JYN1NVW651KCA56C102%26tag%3Dtechkie-20%26linkCode%3Dxm2%26camp%3D2025%26creative%3D165953%26creativeASIN%3D048660456X}
  {\emph {\bibinfo {title} {The Theory of the Properties of Metals and
  Alloys}}}\ (\bibinfo  {publisher} {Dover Publications},\ \bibinfo {year}
  {1958})\BibitemShut {NoStop}%
\end{thebibliography}%

%

\end{document}